\documentclass[twocolumn, osajnl, showpacs, showkeys, 10pt]{revtex4}

\usepackage{graphicx}
\newcommand{\ud}{\mathrm{d}}
\newcommand{\eqn}[1]{Eq.~(\ref{#1})}
\newcommand{\eqs}[1]{Eqs.~(\ref{#1})}
\newcommand{\Fig}[1]{Figure~\ref{#1}}
\newcommand{\Figs}[1]{Figures~\ref{#1}}

\newcommand{\figs}[1]{Figs.~\ref{#1}}
\newcommand{\Ref}[1]{Ref.~\onlinecite{#1}}

\begin{document}
%
\title{Phase Statistics of Soliton}
\author{Keang-Po Ho}
\affiliation{Graduate Institute of Communication Engineering,
National Taiwan University, Taipei 106, Taiwan}
\email{kpho@cc.ee.ntu.edu.tw}

\date{\today}%



\begin{abstract}
The characteristic function of soliton phase jitter is found analytically when the soliton is perturbed by amplifier noise.
In additional to that from amplitude jitter, the nonlinear phase noise due to frequency and timing jitter is also analyzed.
Because the nonlinear phase noise is not Gaussian distributed, the overall phase jitter is also non-Gaussian.
For a fixed mean nonlinear phase shift, the contribution of nonlinear phase noise from frequency and timing jitter decreases with distance and signal-to-noise ratio.
\end{abstract}

\maketitle

\section{INTRODUCTION}

The phase jitter of soliton due to amplifier noise, like Gordon-Haus timing jitter \cite{gordon86}, is usually assumed to be Gaussian distributed \cite{blow92, hanna00, hanna01, iannone}.
When the phase jitter of soliton was studied, the phase jitter variance was given or measured and the statistics of soliton phase is not discussed \cite{blow92, hanna00, hanna99, leclerc98, mckinstrie02}.

For non-soliton systems, the statistics of nonlinear phase noise is found to be non-Gaussian distributed both experimentally \cite{kim03} and theoretically \cite{mecozzi94, ho03asy, ho03pro, ho03sta}.
However, those studies \cite{mecozzi94, kim03, ho03asy, ho03pro, ho03sta} just includes the Gordon-Mollenauer effect \cite{gordon90} that is the nonlinear phase noise induced by the conversion of amplitude to phase jitter due to fiber Kerr effect, mostly self-phase modulation.
Based on the well-developed perturbation theory of soliton \cite{kivshar89, kaup90, georges95, iannone}, phase jitter can also be induced by the interaction of frequency and timing jitter.
In this paper, the statistics of the soliton phase is derived including the contribution of timing and frequency jitter induced nonlinear phase noise.
The characteristic function of soliton phase jitter is derived analytically, to our knowledge, the first time.
The probability density function (p.d.f.) is simply the inverse Fourier transform of the corresponding characteristic function.

Most optical communication systems use the intensity of the optical signal to transmit information. 
Direct-detection differential phase-shift keying (DPSK) signaling has renewed attention recently \cite{gnauck02, miyamoto02, bissessur03, gnauck03, cho03, rasmussen03, zhu03, vareille03, tsuritani03, cai03}, mostly using return-to-zero (RZ) pulses for long-haul transmission and encode information in the phase difference between two consecutive pulses.
To certain extend, a soliton DPSK system may be a good approximation to phase modulated dispersion managed soliton \cite{mckinstrie02} or RZ signal.
With well-developed perturbation theory \cite{kivshar89, kaup90, georges95, iannone}, the distribution of the soliton phase jitter can be derived analytically.

The error probability of DPSK soliton signal was calculated in \Ref{shum97} using the method of Refs. \onlinecite{shum96} and \onlinecite{humblet91} without taking into account the effect of phase jitter.
If the phase jitter is Gaussian distributed, the system can be analyzed by the formulas of \Ref{nicholson84}.
The phase jitter may be indeed Gaussian distributed in certain regimes  around the center of the distribution \cite{hanna00, holzlohner02}, especially if the p.d.f. is plotted in linear scale.
The tail probability less than, for example, $10^{-9}$, is certainly not Gaussian distributed.
As optical communication systems are aimed for very low error probability, a careful study of the statistics of the soliton phase is necessary to characterize the performance of the system.

The remaining parts of this paper are organized as following: Sec. \ref{sec:sto} gives the stochastic equations of the phase jitter according to the first-order soliton perturbation theory; Sec. \ref{sec:cf} derives the characteristic function of soliton phase jitter; Sec. \ref{sec:num} presents the numerical results; and Secs. \ref{sec:dis} and \ref{sec:end} are the discussion and conclusion of the paper, respectively.

\section{STOCHASTIC EQUATIONS FROM SOLITON PERTURBATION}
\label{sec:sto}

From the first-order perturbation theory, with amplifier noise, the soliton parameters evolve according to the following equations  \cite{kivshar89, kaup90, georges95, iannone}

\begin{eqnarray}
\frac{d A}{d \zeta} & = & \Im \left\{ \int  \ud  \tau f_A n(\zeta, \tau) \right\}
 \label{amp} \\
\frac{d \Omega}{ d \zeta} & = & 
           \Re \left\{ \int  \ud  \tau f_\Omega n(\zeta, \tau) \right\} 
  \label{freq}\\
\frac{d T}{ d \zeta} & = & - \Omega + 
           \Im \left\{ \int  \ud  \tau f_T n(\zeta, \tau) \right\} 
  \label{time} \\ 
\frac{d \phi} { d \zeta} & = &
          \frac{1}{2}\left(A^2 - \Omega^2 \right)
	  + T \frac{d \Omega}{d \zeta} + 
	  \Re \left\{ \int  \ud \tau f_\phi n(\zeta, \tau) \right\} 
  \label{phase}
\end{eqnarray} 

\noindent where $\Re\{\ \}$ and $\Im\{\ \}$ denote the real and imaginary part of a complex number, respectively, $n(\zeta, \tau)$ is the amplifier noise with the correlation of

\begin{equation}
E\{n(\zeta_1, \tau_1)n(\zeta_2, \tau_2)\} = \sigma_n^2 \delta(\zeta_1-\zeta_2) \delta(\tau_1 - \tau_2),
\end{equation}

\noindent  $A(\zeta)$, $\Omega(\zeta)$, $T(\zeta)$, and $\phi(\zeta)$ are the amplitude, frequency, timing, and phase parameters of the perturbed soliton of

\begin{eqnarray}
q_0(\tau, \zeta) &=& A(\zeta) \mathrm{sech} \left\{
    A(\zeta)[ \tau - T(\zeta)] \right\} \nonumber \\
& & \times    \exp \left[ - i \Omega(\zeta) \tau + i \phi(\zeta) \right]
\end{eqnarray}

\noindent with initial values of $A(0) = A$ and $\Omega(0) = \phi(0) = T(0) = 0$.
Functions related to soliton parameters are

\begin{eqnarray}
f_A & =& q_0^*, \\
f_\Omega & = & \mathrm{tanh}[A(\tau - T)] q_0^*, \\
f_T & = & \frac{\tau - T}{A} q_0^*, \\
f_\phi & = & -\frac{1}{A}\left\{ 1 - A(\tau -T) \mathrm{tanh}[A(\tau - T)] \right\} q_0^*. 
\end{eqnarray}

From both \eqs{amp} and (\ref{freq}), we get

\begin{eqnarray}
A(\zeta) &=& A + w_A(\zeta) \label{ampsol} \\
\Omega(\zeta) &=& w_\Omega(\zeta)  \label{freqsol}
\end{eqnarray} 

\noindent where $w_A$ and $w_\Omega$ are two independent zero-mean Wiener process with autocorrelation functions of

\begin{eqnarray}
E\{w_A(\zeta_1)w_A(\zeta_2)\} &=& \sigma_A^2 \min(\zeta_1, \zeta_2),\\
E\{w_\Omega(\zeta_1)w_\Omega(\zeta_2)\} &=& \sigma_\Omega^2 \min(\zeta_1, \zeta_2),
\end{eqnarray}
 
\noindent where  $\sigma_A^2 = A \sigma_n^2$ and $\sigma_\Omega^2 = A \sigma_n^2/3$ \cite{georges95, iannone, leclerc98}.
Defined for the amplitude, the signal-to-noise ratio (SNR) as a function of distance is

\begin{equation}
\frac{A^2}{ \sigma_A^2 \zeta} = \frac{A}{ \sigma_n^2 \zeta}.
\label{snr}
\end{equation}

Using \eqs{time} and (\ref{freqsol}), the timing jitter is

\begin{equation}
T(\zeta) = - \int_0^\zeta w_\Omega(\zeta_1) \ud \zeta_1 + w_T(\zeta), \label{timesol}
\end{equation}

\noindent where $w_T$ is a zero-mean Wiener process with autocorrelation function of

\begin{equation}
E\{w_T(\zeta_1)w_T(\zeta_2)\} = \sigma_T^2 \min(\zeta_1, \zeta_2)
\end{equation}

\noindent with \cite{georges95, iannone, leclerc98}

\begin{equation}
\sigma_T^2 = \frac{\pi^2}{12} \frac{ \sigma_n^2}{A}.
\end{equation}

Using \eqs{time}, (\ref{ampsol}), and (\ref{timesol}), the phase jitter is

\begin{eqnarray}
\phi(\zeta) & = & \frac{1}{2} \int_0^\zeta [A + w_A(\zeta_1)]^2 \ud \zeta_1
     - \frac{1}{2} \int_0^\zeta w^2_\Omega(\zeta_1) \ud \zeta_1 \nonumber \\
     & & + \int_0^\zeta \left(-\int_0^{\zeta_1} w_\Omega(\zeta_2) \ud \zeta_2 + w_T(\zeta_1) \right) \ud w_\Omega(\zeta_1) \nonumber \\
& & \qquad + w_\phi(\zeta), 
\label{phasesol}
\end{eqnarray}

\noindent where $w_\phi$ is a zero-mean Wiener process with autocorrelation function of

\begin{equation}
E\{w_\phi(\zeta_1)w_\phi(\zeta_2)\} = \sigma_\phi^2 \min(\zeta_1, \zeta_2)
\end{equation}

\noindent with \cite{georges95, iannone, leclerc98}

\begin{equation}
\sigma_\phi^2 = \frac{\sigma_n^2}{3 A} \left( 1 + \frac{\pi^2}{12} \right).
\end{equation}

The Wiener processes of $w_A, w_\Omega, w_T$, and $w_\phi$ are independent of each other.
The amplitude [\eqn{ampsol}], frequency [\eqn{freqsol}], and timing [\eqn{timesol}]  jitters are all Gaussian distributed.
From \eqn{phasesol}, it is obvious that the phase jitter is not Gaussian distributed.
If \eqn{phase} is linearized or all higher-order terms of \eqn{phasesol} are ignored, the phase jitter is Gaussian distributed and equals to $\phi(\zeta) \approx A \int_0^\zeta w_A(\zeta_1) \ud \zeta_1 + w_\phi(\zeta)$  [\Ref{iannone}].
The characteristic function of the phase jitter \eqn{phasesol} will be derived later in this paper and compared with Gaussian approximation.

\section{CHARACTERISTIC FUNCTIONS OF PHASE JITTER}
\label{sec:cf}

In the phase jitter of \eqn{phasesol}, there are three independent contributions from amplitude jitter (the first term), frequency and timing jitter (the second and third terms), and the projection of amplifier noise to phase jitter $w_\phi$. 
In this section, the characteristic functions of each individual component are derived and the overall characteristic function of phase jitter is the product of the characteristic functions of each independent contribution.

\subsection{Gordon-Mollenauer Effect}

The first term of \eqn{phasesol} is the Gordon-Mollenauer effect \cite{gordon90} of

\begin{equation}
\phi_{\mathrm{GM}}(\zeta) =  \frac{1}{2} \int_0^\zeta [A + w_A(\zeta_1)]^2 \ud \zeta_1,
\label{phigm}
\end{equation}

\noindent induced by the interaction of fiber Kerr effect and amplifier noise, affecting phase-modulated non-return-to-zero (NRZ) and RZ signal \cite{mecozzi94, ho03sta, ho03asy}.

The characteristic function of Gordon-Mollenauer nonlinear phase noise is given by \cite{ho03sta, ho03asy}

\begin{eqnarray}
&&\Psi_{\phi_{\mathrm{GM}}(\zeta)}(\nu)
    =  \sec^{\frac{1}{2}}\left(\zeta \sigma_A \sqrt{j \nu} \right)   \nonumber \\
&&  \qquad \qquad \times \exp \left[\frac{A^2}{2 \sigma_A} \sqrt{j \nu} \mathrm{tan} 
           \left(\zeta \sigma_A \sqrt{j \nu} \right)  \right]. 
\label{cfgm}
\end{eqnarray}

\noindent The above characteristic function \eqn{cfgm} can also be derived from \eqn{quadratic} of the appendix.

The mean and variance of the phase jitter \eqn{phigm} are

\begin{equation}
<\!\!\phi_{\mathrm{GM}}(\zeta)\!\!> = - j \left. \frac{ d}{d \nu}  \Psi_{\phi_{\mathrm{GM}}(\zeta)}(\nu) \right|_{\nu = 0} 
    = \frac{1}{2} A^2 \zeta + \frac{1}{4} \sigma_A^2 \zeta^2,
\label{meangm}
\end{equation}

\noindent and

\begin{eqnarray}
 \sigma_{\phi_{\mathrm{GM}}(\zeta)}^2 & =&  -\left. \frac{ d^2}{d \nu^2}  \Psi_{\phi_{\mathrm{GM}}(\zeta)}(\nu) \right|_{\nu = 0} - <\!\!\phi_{\mathrm{GM}}(\zeta)\!\!>^2  \nonumber \\
	&= &\frac{1}{3} A^2 \sigma_A^2 \zeta^3 + \frac{1}{12} \sigma_A^4 \zeta^4, 
\label{vargm}
\end{eqnarray}

\noindent respectively.
The first term of \eqn{vargm} increases with $\zeta^3$, conforming to that of \Ref{gordon90}. 
Given a large fixed SNR of $A^2/(\sigma^2_A \zeta)$ [\eqn{snr}], the second term of \eqn{vargm} is much smaller than the first term and also increases with $\zeta^3$.
Note that the first term of the mean of \eqn{meangm} is also larger than the second term for large SNR.

The characteristic function of \eqn{cfgm} depends on two parameters: the mean nonlinear phase shift of $A^2 \zeta/2$ and the SNR of \eqn{snr}.
Given a fixed mean nonlinear phase shift of $A^2 \zeta /2$, the shape of the distribution depends only on the SNR \cite{ho03asy}.

Based on \eqn{phasesol}, comparing \eqn{cfgm} with the non-soliton case of \Ref{ho03asy}, the mean and standard deviation of the Gordon-Mollenauer phase noise of soliton are about half of that of non-soliton case with the same amplitude $A$ as the NRZ or RZ level \cite{ho03asy}.

\subsection{Frequency and Timing Effect}

The frequency and timing jitter contributes to phase jitter by 

\begin{eqnarray}
\phi_{\Omega, T}(\zeta) & = &  - \frac{1}{2} \int_0^\zeta w^2_\Omega(\zeta_1) \ud \zeta_1 \nonumber \\
&& \quad - \int_0^\zeta \int_0^{\zeta_1} w_\Omega(\zeta_2) \ud \zeta_2 \ud w_\Omega(\zeta_1) \nonumber \\
&& \qquad + \int_0^\zeta w_T(\zeta_1)  \ud w_\Omega(\zeta_1)
\label{phiot}
\end{eqnarray} 

\noindent as the second and third terms of \eqn{phasesol}.

By changing the order of integration for the second term of \eqn{phiot}, we get

\begin{eqnarray}
\phi_{\Omega, T}(\zeta)  &=&   \frac{1}{2} \int_0^\zeta w^2_\Omega(\zeta_1) \ud \zeta_1  + \int_0^\zeta w_T(\zeta_1)  \ud w_\Omega(\zeta_1) \nonumber \\
& & \qquad - w_\Omega(\zeta) \int_0^{\zeta} w_\Omega(\zeta_1) \ud \zeta_1 .
\end{eqnarray}

\noindent From \eqn{cfVarPhi} of the appendix, the characteristic function of $\phi_{\Omega, T}(\zeta)$ is

\begin{equation}
\Psi_{\phi_{\Omega, T}(\zeta)}(\nu) = \Psi_{\varphi_1, \varphi_2, \varphi_3}\left(\frac{\nu}{2}, \nu, -\nu \right).
\label{cfOT}
\end{equation}

The mean and variance of the phase jitter of \eqn{phiot} are

\begin{equation}
<\!\!\phi_{\Omega, T}(\zeta)\!\!> = - j \left. \frac{ d}{d \nu}  \Psi_{\phi_{\Omega, T}(\zeta)}(\nu) \right|_{\nu = 0} 
    =  - \frac{1}{4} \sigma_\Omega^2 \zeta^2,
\label{meanot}
\end{equation}

\noindent and

\begin{eqnarray}
 \sigma_{\phi_{\Omega, T}(\zeta)}^2 & =& 
   - \left. \frac{ d^2}{d \nu^2}  \Psi_{\phi_{\Omega, T}(\zeta)}(\nu) \right|_{\nu = 0} - <\!\!\phi_{\Omega, T}(\zeta)\!\!>^2  \nonumber \\
	&= &\frac{1}{2} \sigma_\Omega^2 \sigma_T^2 \zeta^2 + \frac{1}{4} \sigma_\Omega^4 \zeta^4,
\label{varot}
\end{eqnarray}

\noindent respectively.

Comparing the means of \eqs{meangm} and (\ref{meanot}), in terms of absolute value,  the mean nonlinear phase shift due to Gordon-Mollenauer effect is much larger than that due to frequency and timing effect.
Comparing the variances of \eqs{vargm} and (\ref{varot}), the variance of nonlinear phase noise due to Gordon-Mollenauer effect is also much larger than that due to frequency and timing effect.

Unlike the Gordon-Mollenauer effect, the characteristic function of \eqn{cfOT}, from the appendix, is not determined only on the SNR and the mean nonlinear phase shift \eqn{meanot}.

\subsection{Linear Phase Noise}

The last term of \eqn{phasesol} gives the linear phase noise of

\begin{equation}
\phi_{\mathrm{LN}}(\zeta) = w_\phi(\zeta)
\end{equation}

\noindent with a characteristic function of

\begin{equation}
\Psi_{\phi_{\mathrm{LN}}(\zeta)}(\nu) 
  = \exp\left(- \frac{1}{2} \sigma_\phi^2 \zeta \nu^2 \right).
\label{cfLN}
\end{equation}

\noindent 
From the characteristic function of \eqn{cfLN}, the linear phase noise depends solely on the SNR [\eqn{snr}].

The characteristic function of the overall phase jitter $\phi(\zeta)$ is the multiplication of the characteristic functions of \eqs{cfgm}, (\ref{cfOT}), and (\ref{cfLN}).

Although the actual mean nonlinear phase shift is 

\begin{equation}
<\!\!\phi(\zeta)\!\!> = <\!\!\phi_{\Omega, T}(\zeta)\!\!> + <\!\!\phi_{\mathrm{GM}}(\zeta)\!\!>,
\end{equation}

\noindent we mostly call $A^2 \zeta/2$ the mean nonlinear phase shift as a good approximation in high SNR.

\section{NUMERICAL RESULTS}
\label{sec:num}

\begin{figure*}
\centerline{
\begin{tabular}{ccc}
    \includegraphics[width = 0.32 \textwidth]{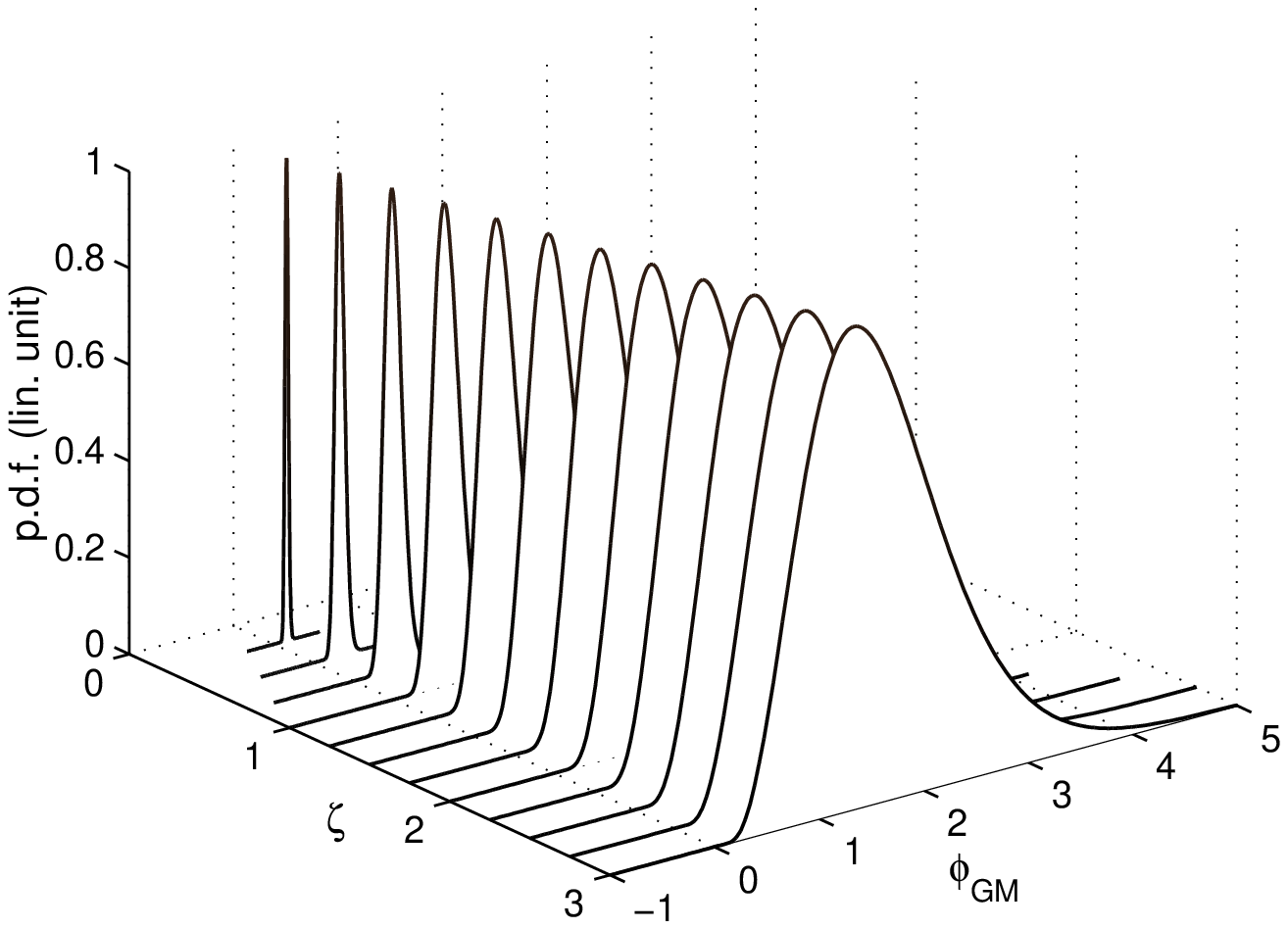} &
    \includegraphics[width = 0.32 \textwidth]{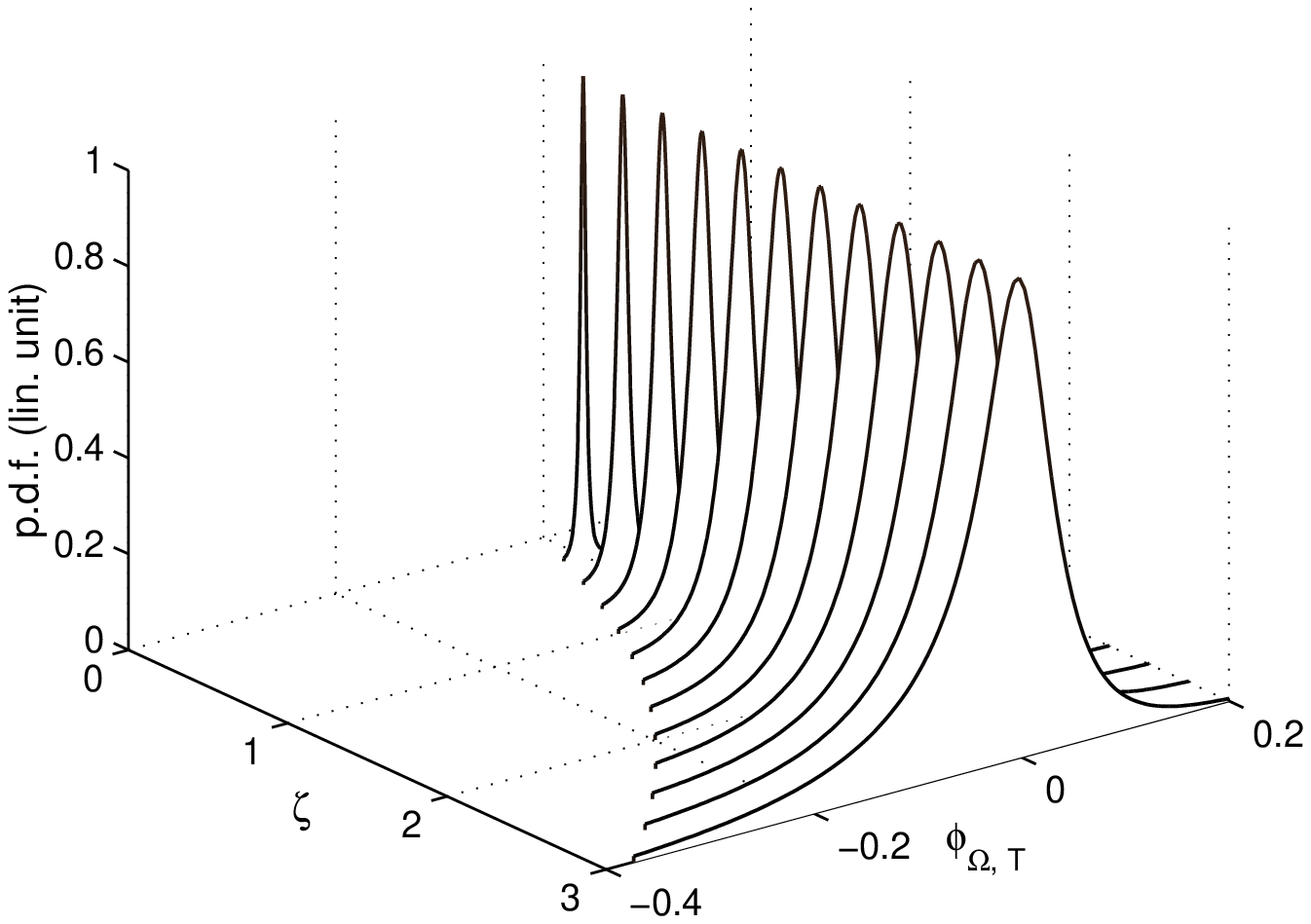} & 
    \includegraphics[width = 0.32 \textwidth]{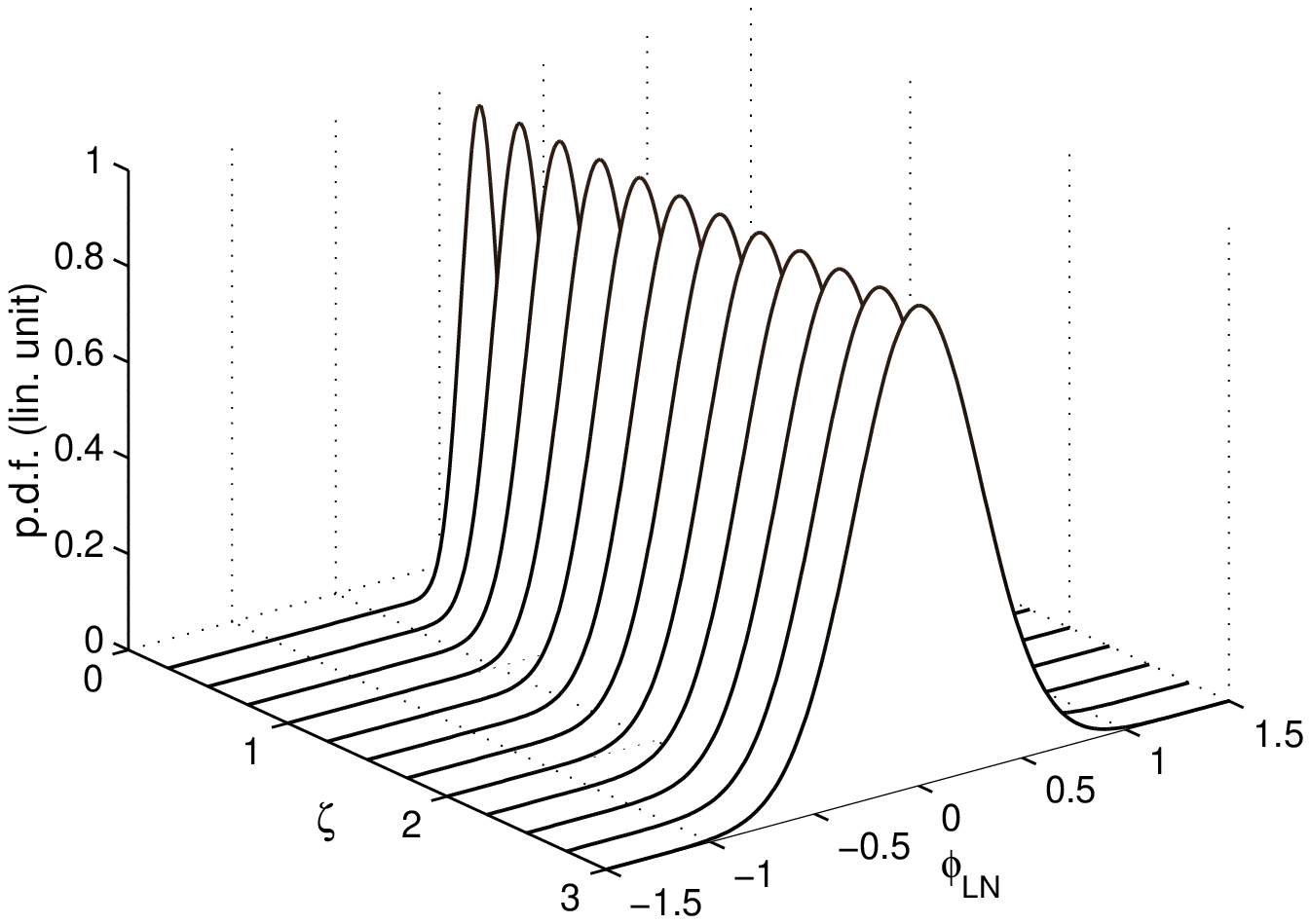} \\
    (a) $\phi_{\mathrm{GM}}(\zeta)$ & (b) $\phi_{\Omega, T}(\zeta)$ & (c) $\phi_{\mathrm{LN}}(\zeta)$ \\
\multicolumn{3}{c}{\includegraphics[width = 0.5 \textwidth]{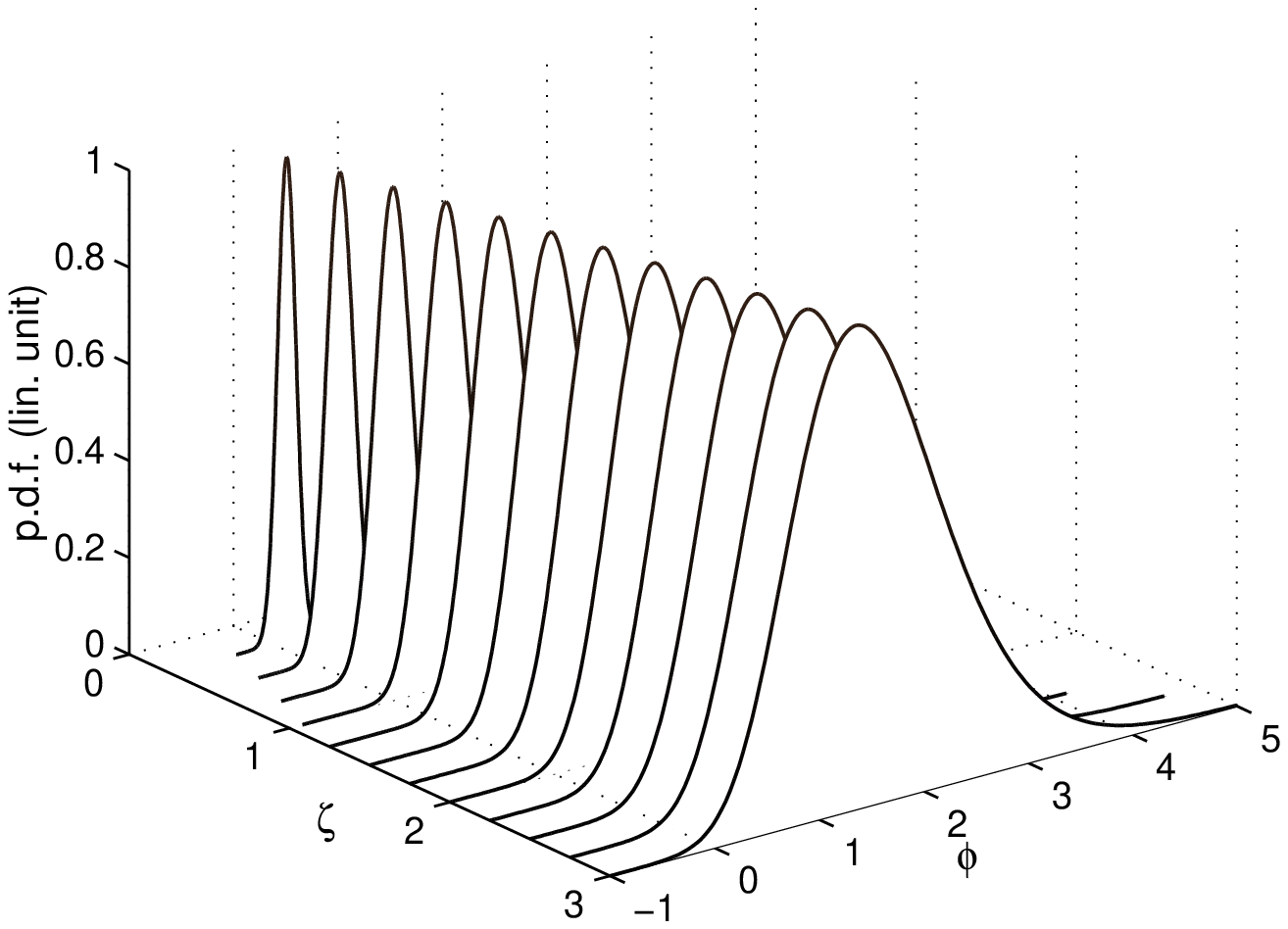}}  \\
 & (d) $\phi(\zeta)$ & 
\end{tabular}
}
\caption{The distributions of soliton phase jitter for difference distance for $A = 1$, $\sigma^2_n = 0.05$. 
The distributions are normalized for a unity peak.
The $x$-axis is not in the same scale.
 }
\label{figpdf}
\end{figure*}

The p.d.f. is the inverse Fourier transform of the corresponding characteristic function.
\Figs{figpdf} show the evolution of the distribution of the phase jitter [\eqn{phasesol}] with distance.
The system parameters are $A = 1$ and $\sigma_n^2 = 0.05$.
Those parameters are chosen for typical distribution of the phase jitter.

\Figs{figpdf}(a), (b), (c) are the distribution of Gordon-Mollenauer nonlinear phase noise [\eqn{cfgm}], frequency and timing nonlinear phase noise [\eqn{cfOT}], and the linear phase noise [\eqn{cfLN}], respectively, as components of the overall phase jitter of \eqn{phasesol}.
\Fig{figpdf}(d) is the distribution of the overall phase jitter \eqn{phasesol}.
The p.d.f.'s in \figs{figpdf} are normalized to a unity peak value for illustration purpose.
The $x$-axis of individual figure of \figs{figpdf} does not have the same scale.
From \figs{figpdf}, the nonlinear phase noises from Gordon-Mollenauer effect and frequency and timing effect are obvious not Gaussian distributed.
With small mean and variance, the nonlinear phase noise from frequency and timing effect has a very long tail.

\begin{figure*}
\centerline{
\begin{tabular}{cc}
    \includegraphics[width = 0.45 \textwidth]{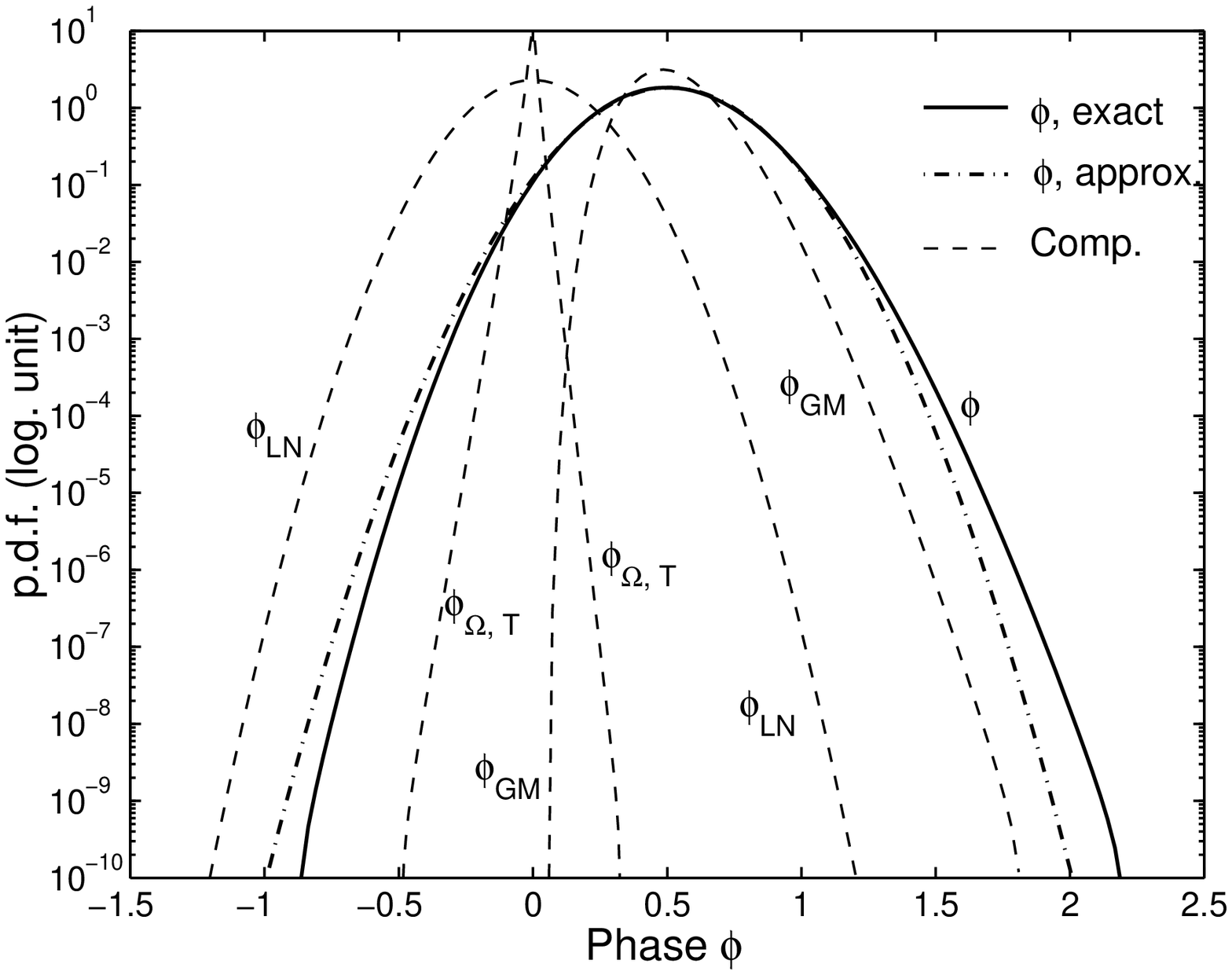} &
    \includegraphics[width = 0.45 \textwidth]{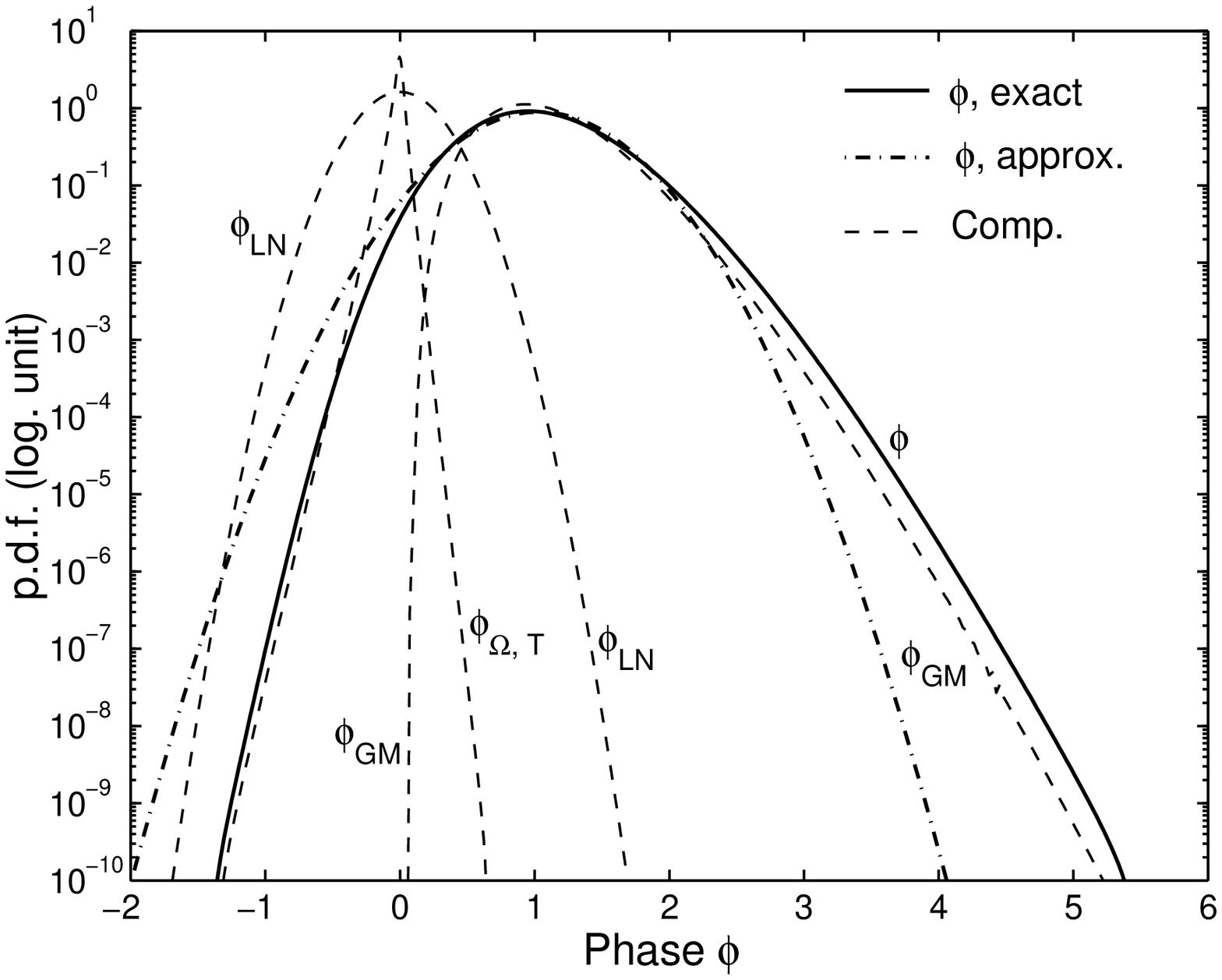} \\
    (a) $\zeta = 1$ & (b) $\zeta =  2$ 
\end{tabular}
}
\caption{The distributions of soliton phase jitter for two distances of (a) $\zeta$ = 1 and (b) $\zeta$ = 2.
 }
\label{figzeta}
\end{figure*} 

\Figs{figzeta} plot the p.d.f.'s of \figs{figpdf} in logarithmic scale for the cases of $\zeta = 1, 2$.
The Gaussian approximation is also plotted in \figs{figzeta} for the overall phase jitter $\phi(\zeta)$.
In both cases of $\zeta = 1, 2$, the Gaussian approximation is not close to the exact p.d.f.'s in the tails.
However, if the p.d.f.'s are plotted in linear scale, Gaussian approximation may be very close to the actual distribution, especially for large phase jitter \cite{holzlohner02}.
The p.d.f's in \figs{figzeta} are not normalized to a unity peak.
  
From both \figs{figpdf} and \ref{figzeta}, the nonlinear phase noises of $\phi_{\mathrm{GM}}$ and $\phi_{\Omega, T}$ are not symmetrical with respect to their corresponding means.
While $\phi_{\mathrm{GM}}$ spreads further to positive phase, $\phi_{\Omega, T}$ spreads further to negative phase.
Plotted in the same scale, the nonlinear phase noise of $\phi_{\mathrm{GM}}$ due to Gordon-Mollenauer effect is much larger than the nonlinear phase noise of  $\phi_{\Omega, T}$ due to frequency and timing effect.

The p.d.f.'s in \figs{figpdf} cannot cover all possible cases. 
While both the Gordon-Mollenauer and linear phase noises depend on the mean nonlinear phase shift $A^2 \zeta/2$ and SNR, the nonlinear phase noise induced by frequency and timing effect does not have a simple scaled relationship.

\begin{figure*}
\centerline{
\begin{tabular}{cc}
    \includegraphics[width = 0.45 \textwidth]{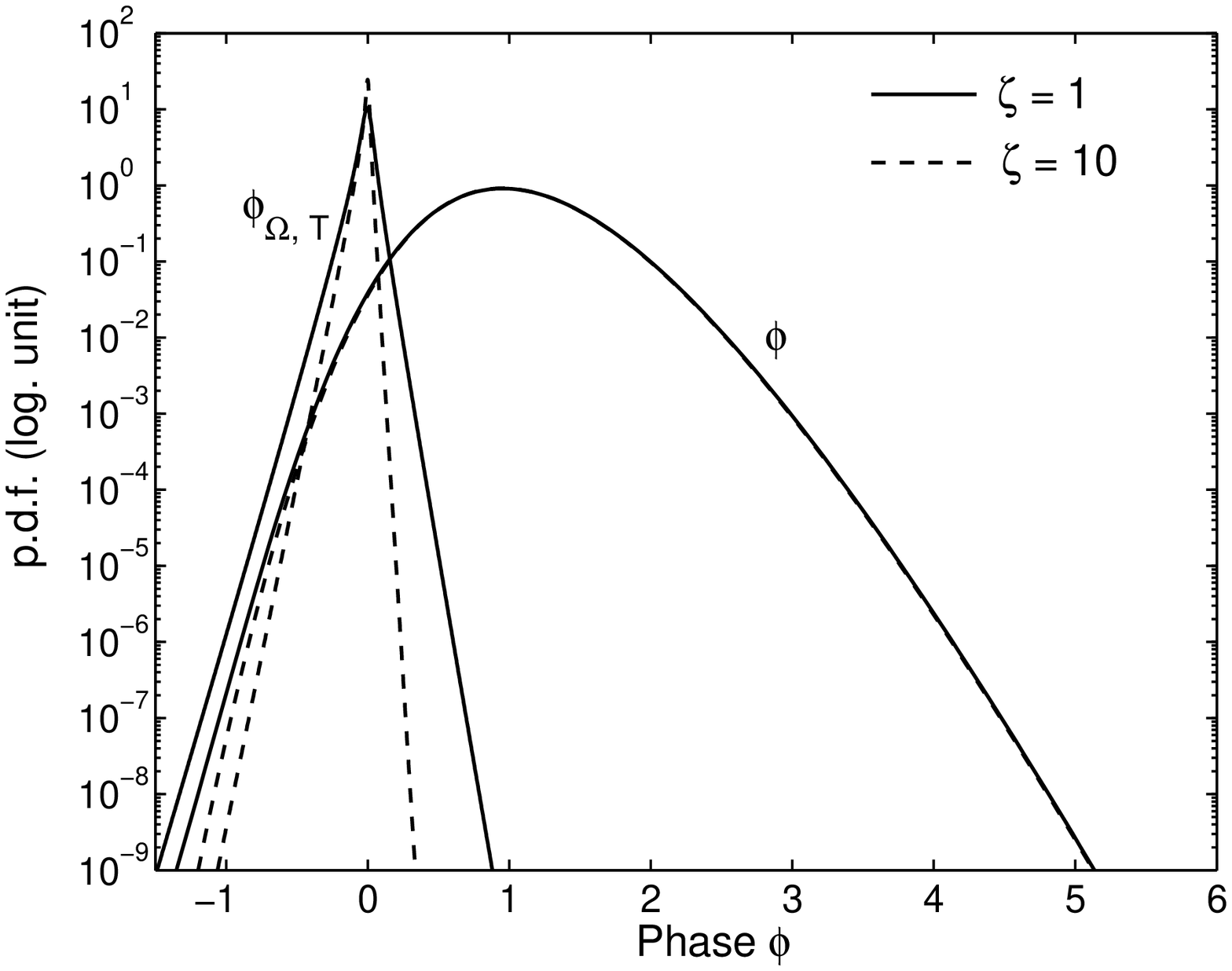} &
    \includegraphics[width = 0.45 \textwidth]{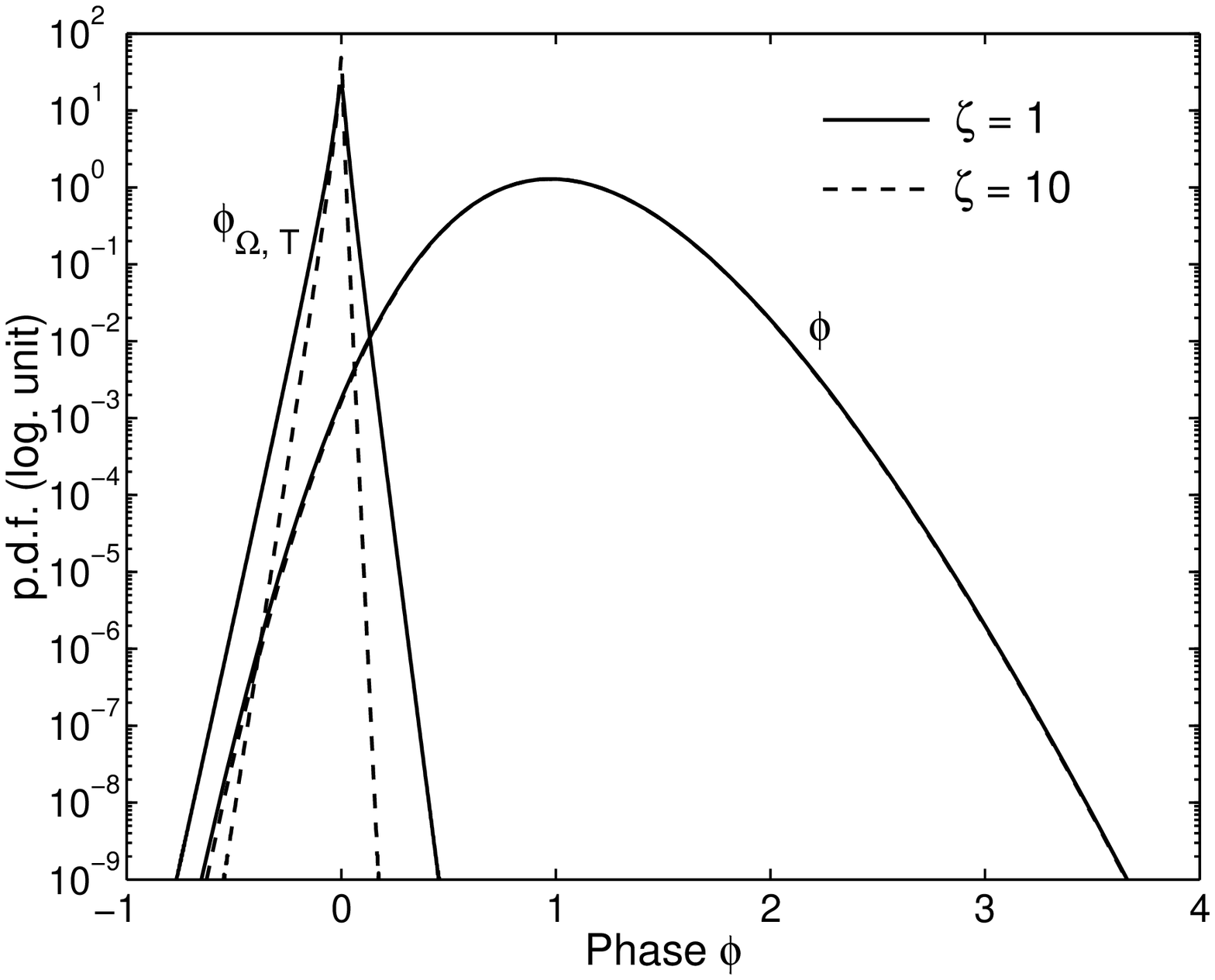} \\
    (a) SNR of 10 & (b) SNR of 20 
\end{tabular}
}
\caption{The distributions of soliton phase jitter for SNR of (a) 10 and (b) 20.
 }
\label{figsnr}
\end{figure*}

For a mean nonlinear phase shift of $\frac{1}{2} A^2 \zeta = 1$ rad \cite{gordon90}, \Figs{figsnr} plot the distribution of the overall phase jitter [\eqn{phasesol}] for a SNR of 10 and 20 for $\zeta = 1, 10$.
After a scale factor, the distributions of both Gordon-Mollenauer and linear phase noise are the same as that in \figs{figzeta}.
In additional to the overall phase jitter, \Figs{figsnr} also plot the distribution of the nonlinear phase noise from frequency and timing effect of $\phi_{\Omega, T}$.

For a fixed mean nonlinear phase shift and SNR, from \figs{figsnr}, the nonlinear phase noise from frequency and timing effect of $\phi_{\Omega, T}(\zeta)$ has less effect to the overall phase jitter for long distance than short distance.
\Figs{figpdf} are plotted for short distance of $\zeta \leq 3$ to show the contribution of frequency and timing jitter to nonlinear phase noise.
The effect of  $\phi_{\Omega, T}(\zeta)$ is smaller for large SNR of 20 than small SNR of 10.
The main contribution to the overall phase jitter is always the Gordon-Mollenauer effect and the linear phase noise.

\section{DISCUSSION}
\label{sec:dis}

The phase jitter of \eqn{phasesol} is derived based on the first-order perturbation theory \cite{kivshar89, kaup90, georges95, iannone} of \eqs{amp} to (\ref{phase}). 
The non-Gaussian distribution is induced by the higher-order terms of \eqn{phasesol} or the nonlinear terms of \eqn{phase}.
Second- and higher-order soliton perturbation \cite{kaup91, haus97} may give further non-Gaussian characteristic to the phase jitter.
Currently, there is no comparison between contributions of the higher-order terms of \eqn{phase} and higher-order soliton perturbation.

In this paper, like almost all other literatures \cite{gordon86, blow92, iannone, leclerc98, mckinstrie02, gordon90, kivshar89, kaup90, georges95}, the impact of amplitude jitter to the noise variances of $\sigma^2_A$, $\sigma^2_\Omega$, $\sigma^2_T$, and $\sigma^2_\phi$ is ignored.
The noise variances of $\sigma^2_A$, $\sigma^2_\Omega$, $\sigma^2_T$, and $\sigma^2_\phi$ are assumed independent of distance.
If the amplitude noise variance is $\sigma_A^2 = A(\zeta) \sigma_n^2$ with dependence on the instantaneous amplitude jitter, amplitude, frequency, and timing jitters are all non-Gaussian distributed \cite{ho03non}.
As an example, amplitude jitter is non-central chi-square distributed \cite{moore03, ho03non}. 
However, the statistics of phase jitter [\eqn{phasesol}] does not have a simple analytical solution when the noise variance depends on amplitude jitter.
With a high SNR, the amplitude jitter is always much smaller than the amplitude $A(0) = A$. 
Even in high SNR, the phase jitter is non-Gaussian based on \eqn{phasesol}.
 
\section{CONCLUSION}
\label{sec:end}

Based on the first-order soliton perturbation theory, the distribution of soliton phase jitter due to amplifier noise is derived analytically the first time.
In additional to the main contribution of Gordon-Mollenauer effect, the nonlinear phase noise due to frequency and timing jitter is also considered.
Induced by Gordon-Mollenauer effect or frequency and timing jitter, the nonlinear phase noises are not Gaussian distributed, neither does the overall phase jitter.
For a fixed mean nonlinear phase shift, the contribution of nonlinear phase noise from frequency and timing jitter decreases with distance and SNR.

\appendix
\section{}
\setcounter{equation}{0}
\renewcommand{\theequation}{A.\arabic{equation}}

Here, we find the joint characteristic function of 

\begin{eqnarray}
\varphi_1 &=& \int_0^\zeta w^2_\Omega(\zeta_1) \ud \zeta_1, \\ 
\varphi_2 &=& \int_0^\zeta w_T(\zeta_1) \ud w_\Omega(\zeta_1), \\
\varphi_3 &=&  w_\Omega(\zeta) \int_0^{\zeta} w_\Omega(\zeta_1) \ud \zeta_1.
\end{eqnarray}

\begin{widetext}

By changing the integration order, we get

\begin{equation}
\varphi_2 = \int_0^\zeta \int_0^{\zeta_1} \ud w_T(\zeta_2) \ud w_\Omega(\zeta_1)
          = \int_0^\zeta \left[w_\Omega(\zeta) - w_\Omega(\zeta_2) \right] \ud w_T(\zeta_2).
\end{equation}

The joint characteristic function of $\varphi_1$, $\varphi_2$, and $\varphi_3$ is

\begin{equation}
\Psi_{\varphi_1, \varphi_2, \varphi_3}(\nu_1, \nu_2, \nu_3)
    = E\left\{\exp(j \nu_1 \varphi_1 + j \nu_2 \varphi_2 + j \nu_3 \varphi_3)\right\}.
\label{cfVarphi_12}
\end{equation}

\noindent Similar to option pricing with stochastic volatility \cite{stein91}, the expectation of \eqn{cfVarphi_12} can be evaluated in two steps, first over $w_T$ and than $w_\Omega$.
In the average over $w_T$, it is obvious that $\varphi_2$ is a zero-mean Gaussian random variable with a variance of $\sigma^2_T \int_0^\zeta \left[w_\Omega(\zeta) - w_\Omega(\zeta_1) \right]^2 \ud \zeta_1$, we get

\begin{eqnarray}
\Psi_{\varphi_1, \varphi_2, \varphi_3}(\nu_1, \nu_2, \nu_3) &= &     E\Bigg\{ -\frac{ \sigma_T^2 \nu_2^2 }{2} \int_0^\zeta \left[w_\Omega(\zeta) - w_\Omega(\zeta_1) \right]^2 \ud \zeta_1 \nonumber \\  
& &     \qquad    +  j \nu_1 \int_0^\zeta w^2_\Omega(\zeta_1) \ud \zeta_1 + j \nu_3 w_\Omega(\zeta) \int_0^{\zeta} w_\Omega(\zeta_1) \ud \zeta_1 \Bigg\}  \nonumber \\
 & = & E\Bigg\{ -\frac{ \sigma_T^2 \nu_2^2 \zeta}{2} w_\Omega^2(\zeta) 
       + (j \nu_3 + \sigma_T^2 \nu_2^2) w_\Omega(\zeta)\int_0^\zeta w_\Omega(\zeta_1) \ud \zeta_1  \nonumber \\
& & \qquad +   \left(j \nu_1 -\frac{ \sigma_T^2 \nu_2^2 }{2} \right)  \int_0^\zeta w^2_\Omega(\zeta_1) \ud \zeta_1\Bigg\}.
\label{psi12}	 
\end{eqnarray}

First of all, we have \cite{cameron45, mecozzi94, mecozzi94a, ho03sta}

\begin{eqnarray}
\lefteqn{ E\left\{ j \omega_1 w_\Omega(\zeta) + j \omega_2 \int_0^\zeta w_\Omega(\zeta_1) \ud \zeta_1 + \frac{j \omega_3}{2}  \int_0^\zeta w^2_\Omega(\zeta_1) \ud \zeta_1 \right\} }  \nonumber \\
& = & \sec^{\frac{1}{2}} \left( \sqrt{j \omega_3} \sigma_\Omega \zeta \right) \exp\Bigg\{
      -\frac{1}{2} \left( \omega_1^2 \sigma^2_\Omega + \frac{\omega_2^2}{j \omega_3 }  \right) \frac{\tan \left( \sqrt{j \omega_3} \sigma_\Omega \zeta \right)}{\sqrt{j \omega_3} \sigma_\Omega } \nonumber  \\  
& & \qquad \qquad \qquad \qquad \qquad \quad + j \frac{\omega_1 \omega_2}{\omega_3} \left[ \sec\left( \sqrt{j \omega_3} \sigma_\Omega \zeta \right) -1 \right] - j \frac{ \omega_2^2 \zeta}{2 \omega_3}
    \Bigg\} \nonumber \\
& = &  \sec^{\frac{1}{2}} \left( \sqrt{j \omega_3} \sigma_\Omega \zeta \right) 
       \exp \left[
          - \frac{1}{2} \mbox{\boldmath $\omega$}^{T}_{1,2} \mathcal{C}(j\omega_3)  \mbox{\boldmath $\omega$}_{1,2} \right],
 \label{quadratic}
\end{eqnarray}

\noindent where $\mbox{\boldmath $\omega$}_{1,2} = (\omega_1, \omega_2)^T$ and 

\begin{equation}
 \mathcal{C}(j\omega_3) = \left[
  \begin{array}{cc}
    \displaystyle \frac{\sigma_\Omega \tan \left( \sqrt{j \omega_3} \sigma_\Omega \zeta \right)}{\sqrt{j \omega_3} } & 
         \displaystyle  \frac{1}{j \omega_3} \left[ \sec \left( \sqrt{j \omega_3} \sigma_\Omega \zeta \right) -1 \right] \\
     \displaystyle \frac{1}{j \omega_3} \left[ \sec \left( \sqrt{j \omega_3} \sigma_\Omega \zeta \right) -1 \right] & 
       \displaystyle \frac{1}{j \omega_3} \left[
   \frac{\tan \left( \sqrt{j \omega_3} \sigma_\Omega \zeta \right)}{\sqrt{j \omega_3}  \sigma_\Omega} - \zeta \right]
   \end{array}
\right].
\end{equation}

As a verification, if $\omega_3$ approaches zero, the covariance matrix is

\begin{equation}
\lim_{\omega_3 \rightarrow 0} C(j \omega_3) = \sigma_\Omega^2 \left[
 \begin{array}{cc}
   \zeta & \frac{1}{2} \zeta^2 \\
 \frac{1}{2}\zeta^2 & \frac{1}{3} \zeta^3
\end{array}
\right],
\label{covlim}
\end{equation}

\noindent that is the covariance matrix of the vector of

\begin{equation}
\mathbf{w}_\zeta = \left(w_\Omega(\zeta), \int_0^\zeta w_\Omega(\zeta_1) \ud \zeta_1\right)^T
\end{equation}

\noindent without any dependence on the random variable $\varphi_1$.  Note that the equation corresponding to \eqn{quadratic} in Refs. [\onlinecite{mecozzi94, mecozzi94a}] does not have the limit of \eqn{covlim}.

The characteristic function of \eqn{quadratic} is that of a correlated two-dimensional Gaussian random variable of $\mathbf{w}_\zeta$ with dependence to $\varphi_1$. The first two terms of \eqn{psi12} is a quadratic (or bilinear) function of $\mathbf{w}_\zeta$, i.e., $\frac{1}{2} \mathbf{w}^T_\zeta \mathcal{M}(j \nu_2, j \nu_3) \mathbf{w}_\zeta$, where

\begin{equation}
\mathcal{M}(j \nu_2, j \nu_3) = \left[
    \begin{array}{cc}
      -\sigma_T^2 \nu^2_2 \zeta & j \nu_3 + \sigma_T^2 \nu^2_2 \\
     j \nu_3 + \sigma_T^2 \nu^2_2 & 0
   \end{array}
\right].
\end{equation}

\noindent The characteristic function of the quadratic function of zero-mean Gaussian random variables is $\det [ \mathcal{I} - \mathcal{C} \mathcal{M}]^{-\frac{1}{2}}$ [\Ref{ho03pro}], where $\det[\ ]$ denotes the determinant of a matrix. 

The joint characteristic function is

\begin{equation}
 \Psi_{\varphi_1, \varphi_2, \varphi_3}(\nu_1, \nu_2, \nu_3) = \frac { 
\sec^{\frac{1}{2}} \left( \sqrt{2 j\nu_1 - \sigma_T^2 \nu_2^2} \sigma_\Omega \zeta \right) }
    {\det\left[\mathcal{I} - \mathcal{C}\left(2 j\nu_1 - \sigma_T^2 \nu_2^2 \right) \mathcal{M}(j \nu_2, j \nu_3) \right]^{\frac{1}{2}}}. 
\label{cfVarPhi}
\end{equation}

\end{widetext}

\noindent where $\mathcal{I}$ is the identity matrix.
The substitute of $j \omega_3$ by $2 j \nu_1 - \sigma_T^2 \nu_2^2$ is obvious by comparing \eqs{psi12} and (\ref{quadratic}).

We can get \cite{cameron45, ho03asy}  

\begin{equation}
\Psi_{\varphi_1}(\nu_1) = \sec^{\frac{1}{2}} \left( \sqrt{2j \nu_1} \sigma_\Omega \zeta \right)
\end{equation}

\noindent and \cite{stein91}

\begin{equation}
\Psi_{\varphi_2}(\nu_2) = \mathrm{sech} ^{\frac{1}{2}}\left( \sigma_T \sigma_\Omega \zeta \nu_2 \right),
\end{equation}

\noindent respectively. 
We can also get 

\begin{equation}
\Psi_{\varphi_3}(\nu_3) =
  \left[ 1 - j \nu_3 \sigma_\Omega^2 \zeta^2 + \frac{1}{12} \nu_3^2 \sigma_\Omega^4 \zeta^4 \right]^{-\frac{1}{2}}.
\end{equation}

While both random variables $\varphi_1$ and $\varphi_2$ determine by $\sigma_\Omega \zeta$, the random variable of $\varphi_2$ determines by $\sigma_T \sigma_\Omega \zeta$.


\end{document}